\documentclass[a4paper,twocolumn,showpacs,prl,groupedaddress,floatfix,citeautoscript]{revtex4-1}
\usepackage{graphicx}
\usepackage{natbib}
\newcommand{\be}{\begin{equation}}
\newcommand{\ee}{\end{equation}}

\begin{document}
\title{Donor and Acceptor Levels in Semiconducting Transition Metal Dichalcogenides}
\author{A. Carvalho$^{1}$, A.~H. Castro Neto$^{1}$}

\affiliation{$^{1}$Graphene Research Center, National University of Singapore, 6
Science Drive 2, Singapore 117546}

\email{physdca@nus.sg.edu}

\date{\today}
\begin{abstract}
{
Density functional theory calculations are used to show that 
it is possible to dope semiconducting transition metal dichalcogenides (TMD) such as  MoS$_2$ and WS$_2$ with electrons and/or holes either by chemical substitution or by adsorption on the sulfur layer. Notably, the activation energies of Lithium and Phosphorus, a shallow donor and a shallow acceptor, respectively, are smaller than 0.1~eV. Substitutional halogens are also proposed as alternative donors adequate for different temperature regimes.
All dopants proposed result in very little lattice relaxation and, hence, are expected to lead to minor scattering of the charge carriers. Doped MoS$_2$ and WS$_2$ monolayers are extrinsic in a much wider temperature range than 3D semiconductors, making them superior for high temperature electronic and optoelectronic applications.
}
\end{abstract}

%\section{Introduction}

\date{\today}
\maketitle

%Relevance of MoS2

Advances in the fabrication and characterization of two-dimensional (2D) dichalcogenide semiconductors have 
reshaped the concept of thin transistor gate.\cite{eda-review,wang-science-review}
Unlike thin fully-depleted silicon channels, physically limited by the oxide interface,
single layer metal dichalcogenides are intrinsically 2D and, therefore, have no surface dangling bonds. 
The monolayer thickness is constant, and the scale of the variations of the 
electrostatic potential profile perpendicular
to the plane is only limited by the extent of the electronic wavefunctions. 
Hence, TMD can in principle be considered immune to channel thickness modulation close to the drain.

Building on these fundamental advantages, 
numerous field-effect transistor (FET) designs employing MoS$_2$ or WS$_2$ channels have been proposed.
These range from 2D adaptations of the traditional FET structure, 
where the 2D semiconductor is separated by a dielectric layer from a top gate electrode, to dual-gate heterolayer devices where the transition metal dichalcogenide is 
straddled between two graphene sheets\cite{wang-science-review}.
Such FETs can be integrated into logic inversion circuits,
providing the building blocks for all logical operations \cite{radisavljevic}.

%importance of chemical doping

However, at present the success of TMD in electronics is limited by the difficulty in achieving high carrier concentrations and, by consequence, high electronic mobilities (current values range around 100 cm$^2$/V.s)\cite{radisavljevicnano}.
In the absence of a chemical doping technology, 
the control of the carrier concentration relies solely on the application
of a gate voltage perpendicular to the layer,
which shifts the Fermi level position rendering the material $n$- or $p$-type\cite{ambipolar transistor paper}.
But in practice the gate voltage drop across the insulator cannot exceed its electric breakdown limit 
(about 1 V/nm for SiO$_2$, or lower for high-$\kappa$ dielectrics\cite{sire-APL-91-242905}).
A work-around demonstrated in graphene consists on gating with ferroelectric polymers\cite{guangxin-ACSN-6-3935}, 
although at the expense of the thermal stability and switching time.

%introduce our idea

In this article we use first-principles calculations to show that
MoS$_2$ and WS$_2$ can be doped
both $n$- and $p$-type using substitutional impurities.
This grants transitional metal dichalcogenides an advantage
over other chalcogenide semiconductor families where doping asymmetries are 
notorious:
ZnS can be doped $n$-type but not $p$-type,
while chalcopyrite CuInTe$_2$ and CuGaSe$_2$ can be $p$-type doped but not $n$-type doped\cite{zhao-APL-85-5860},
and SnTe has not yet been doped $n$-type\cite{singh}.
In transition metal dichalcogenides, 
even though chemical doping is mostly unexplored, 
there have already been some experimental reports of successful chemical doping\cite{fang-NL-13-1991,du-EDL-IEEE},as well as some electronic structure calculations for impurities\cite{komsa-PRL-109-035503,ataca}.

Further, we find both $n$- and $p$-type dopants substituting {\it in the S lattice site} 
or adsorbed {\it on top} of the S layer.
Leaving the transition metal layer nearly undisturbed, 
these substitutions promise less scattering to
charge carriers at the Mo-derived states at the bottom of the conduction band (CBM)
or at the top of the valence band (VBM).

%superiority of doped 2D materials

\begin{figure}[p]
\includegraphics[width=8.5cm]{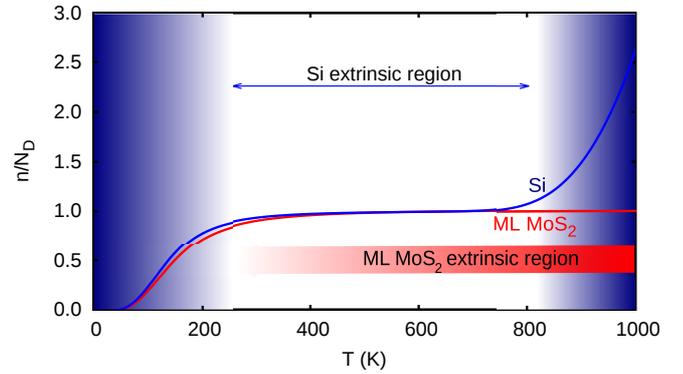}
\caption{
Electron density in $n$-type monolayer MoS$_2$ and Si, 
with concentration $N_D=10^{18}$~cm$^{-3}$
of donors with ionization energy $E_c-E_I=0.045$~eV.
An effective thickness of 6.46~\AA was used for MoS$_2$.
}
\label{fig:plot}
\end{figure}

Having established that doping is possible, 
it follows that 2D doped semiconductors stand out as superior to 3D semiconductors for high temperature applications because of fact that the electronic density of states, $N(E)$, close to the edge of the valence and conduction bands is, unlike the 3D case, energy independent. It is well-known that the intrinsic carrier concentration of a semiconductor is given by: 
\be n_i (T) =\sqrt{N_c(T) N_v(T)}\exp(-E_g/(2kT)), \ee
where $E_g$ is the gap energy, and $N_{c(v)}$ depend on $N(E)$ (and hence the dimensionality) 
of the semicondutor. In 2D we have:
\be N_{c (v)}=\frac{M_{c (v)} m_{e (h)} \ln 2}{\pi\hbar^2} kT,\ee
$M_{c (v)}$ is the degeneracy of the conduction (valence) band, $m_{e (h)}$ is the effective mass of the conduction (valence) band electrons, and $T$ the temperature ($k$ and  $\hbar$ are the Boltzmann and Planck's constants, respectively). Hence, in 2D we have $n_{i,2D}(T) \propto T$ which should be contrasted the 3D counterpart where $n_{i,3D} \propto T^{3/2}$. Figure~\ref{fig:plot} illustrates the relevance of the temperature dependence of the density of conduction electrons $n(T)$,
by comparing the carrier density for $n$-type monolayer MoS$_2$ and
Si, doped with the same dopant concentration and dopant activation energy, 
as a function of temperature.
While Si leaves the extrinsic regime (that is, the region of temperatures where $n_i(T)$ becomes temperature independent) above 800~K, in MoS$_2$ the $n_(T)$ curve is flat beyond 1000~K.
The temperature stability of $n_i$ ultimately reflects on 
transistor characteristics, in particular the gate voltage threshold.

%\section{Method}

We studied donor and acceptor impurities using first-principles calculations.
These were based on density functional theory (DFT), as implemented in the {\sc Quantum ESPRESSO} code.\cite{Giannozzi2009}.
Geometry optimizations and total energy calculations are non-relativistic.
A fully relativistic formalism was used for the bandstructure calculations (see Supplementary Information). 
The exchange correlation energy was described by the generalized gradient approximation (GGA), in the scheme
proposed by Perdew-Burke-Ernzerhof\cite{Perdew1996} (PBE).
The Kohn-Sham bandgaps obtained in the non-relativistic calculations are respectively 
1.65 and 1.77~eV for MoS$_2$ and WS$_2$.
With spin orbit coupling, these values become 1.55 and 1.51~eV, respectively.
We thus find that the GGA is a good approach for bandstructure calculations of these materials, 
and further exchange and correlation effects are likely to produce, in first approximation,
only a rigid shift of the conduction band\cite{komsa-PRB-86-241201}.
The energy cutoff used was 50~Ry. 
Further details of the calculation method can be found in Ref.~\onlinecite{carvalho-PRB} 

The supercell consisted of 4$\times4$ unit cells of the single layer material,
separated by a vacuum spacing with the thickness of two times the supercell lattice parameter. 
For charged supercells, the electrostatic correction of Komsa and Pasquarello was implemented.\cite{komsa-PRL-110-095505,note}
The Brillouin-zone (BZ) was sampled using a 4$\times$4$\times$1 Monkhorst-Pack grid.\cite{Monkhorst1968}

%\section{Defect Structure}

\begin{figure}[p]
\includegraphics[width=8.5cm]{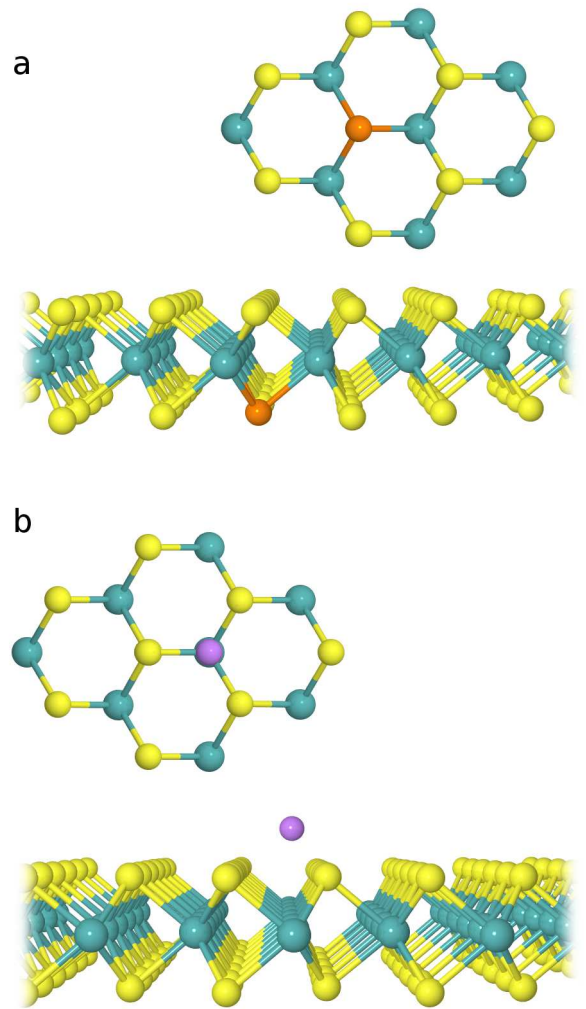}
\caption{
Top: geometry of a distorted substitutional defect (Br$_{\rm S}$),
 in top and side view.
Bottom: geometry of Li adsorbed at the S3 position.
 TM and S atoms are represented as gray and white spheres, respectively.
 The broken bond is represented in dashed line.
}
\label{fig:struc}
\end{figure}

We have considered five dopants: Si, P, Li, Br and Cl.
Any of these can occupy substitutional positions or be adsorbed on the S layer.
The point symmetry of the S site is $C_{3v}$. 
When replaced by P or Si, the resulting defect keeps the trigonal symmetry and
there is little associated lattice distortion.
In the case of neutral Cl$\rm _S$ and Br$\rm _S$ however,
the lowest energy configuration is a $C_s$ geometry where
the neutral Cl$\rm _S$ and Br$\rm _S$ defects are
displaced in the vertical plane, loosening one of the Cl/Br-Mo/W bonds (Fig.~\ref{fig:struc}-a). 
This unusual configuration results from the fact that the halogen partially donates the
unpaired electron to the Mo/W $d$ orbitals, whereas in most molecules Cl and Br receive an electron instead.

Li is most stable at the S3 position\cite{ataca}, shown in Fig.~\ref{fig:struc}-b,
outside the S layer but on the top of a Mo atom.
As for the adsorbed atoms, P, Si, Cl and Br take the S4 configuration as described in Ref.~\onlinecite{ataca},
on top of an S atom. 

%\section{Stability}

\begin{figure}[p]
\includegraphics[scale=1]{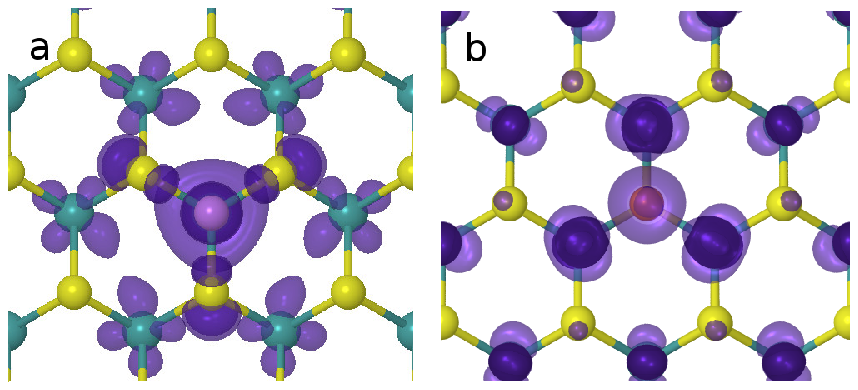}
\caption{
Isosurfaces of the unpaired electron state of Li$\rm _{ad}$ (a)
and of the unpaired hole state of P$\rm _S$ (b), in MoS$_2$,
as generated by fully relativistic calculations.
The former is a donor, whereas the latter is an acceptor.
The square of the wavefunction is represented.
W and S are represented by cyan and yellow spheres, respectively.
}
\label{fig:denpot}
\end{figure}

%\subsection{Formation energy}

A requirement for successful doping is that the impurity must be stable
at the lattice position where it is active,
and comparatively unstable or electrically neutral at the competing positions.
The equilibrium concentration $[D]$ of a defect form $D$ can be related to
the defect formation energy $\Delta G_D$,
\be [D]=gN_D\exp\left(-\frac{\Delta G_D}{kT}\right),\ee
where $N_D$ is the number of sites available to the defect.

Since our calculations are for $T=0$ entropy terms can be neglected,
and the formation energy of the defect can be obtained from the total energies,
\be\Delta G_D\simeq E_f(D)=E(D)-\sum_in_i\mu_i+q\mu_e,\label{eq:Ef}\ee
where $E_f(D)$ is the free energy of the system containing the defect,
$n_i$ is the number of atoms of species $i$ that it contains and
$\mu_i$ is the respective chemical potential. 
Additionally, the formation energy of a charged system, in charge state $q$,
depends on the chemical potential of the electrons ($\mu_e$).

%chemical potentials
The chemical potentials are defined by the experimental growth conditions,
which can range from metal-rich to sulfur-rich.
Bulk MoS$_2$ and WS$_2$ are often sulfur deficient,\cite{sulphur-defMoS2,sulphur-defWS2}
even though sulfur excess has been reported as well\cite{sulphur-excess} 
Here, we will assume $\{\mu_{\rm Mo/W}$,$\mu_{\rm S}\}$ are in the
metal-rich extreme ie.  the system is in equilibrium with a hypothetical reservoir of metallic Mo (or W).
The chemical potentials for the impurities are taken to be the total energy of
the respective isolated atoms, so that $E_f(D_{\rm ad})$ is by definition the adsorption energy.

The calculated formation energies are given in Table~\ref{tab:Esub}.
In the sulfur-poor limit, for both host materials, substitutional Si and P  bind strongly to the lattice,
and are more stable in the substitutional position. 
Li, on the contrary, is most stable at a surface adsorbed position.

Br and Cl have comparable formation energies in both forms.
The energy difference between adsorption
and substitution at the S site is linear on the chemical potential of sulfur,
and independent on the chemical potential of the impurity itself:
\be E_f(D_S)-E_f(D_{\rm ad})=E(D_S)-E(D_{\rm ad})+\mu_{\rm S}.\ee
Thus, it is in principle possible to control the relative populations of Cl or Br
in different sites by changing the sulfur abundance.

Another way to enhance the incorporation ratio of Br and Cl at S sites
by using material that has sulfur vacancies {\rm a priori} (for example pre-irradiated material).
The capture of an impurity atom adsorbed at the layer surface by a sulfur vacancy,
\be {\rm V_S}+X_{\rm ad} \rightarrow  X_{\rm S},\ee
where $\rm {V_S}$ is the sulfur vacancy and $X_{\rm ad}$ is the adsorbed atom
is isoenthalpic for Br and Cl.
Furthermore, for Cl the respective energy gain is actually greater than
the formation energy of the vacancy (1.3 and 1.7~eV in sulfur-poor MoS$_2$ and WS$_2$, 
respectively).

%\section{Electronic levels}

We have so far considered the stability of the neutral defects.
Now the most important requirement for a dopant is that its ionization energy $E^D_I$ 
is not greater than a few $kT$.
The thermodynamic transition level $E_D(q/q+1)$ can be defined
as the value of the Fermi level for which charge states $q$ and $q+1$
of the defect $D$ have the same formation energy.
The position of the  $E_D(q/q+1)$ level relative to the valence band top $E_v$
can be found from the formation energies (see Eq.~\ref{eq:Ef})\cite{footnote}
\begin{eqnarray} E_D(q/q+1)= E_f[X^q]-E_f[X^{q+1}]-E_v.\label{eq:levels}\end{eqnarray}
%corrections
Thus for acceptors $E_D(0/+)\equiv E^D_I$ and for donors $E_D(-/0)\equiv E_g-E^D_I$.

%definition of marker emethod
For comparison, we have also calculated the same
defect levels using the marker method (MM). 
In this method, the ionisation energies/electron affinities of defective supercells
are compared with those of the pristine supercell,\cite{coutinho-PRB-67-035205} 
and the spurious electrostatic interactions are partially canceled.
There is good agreement between the levels calculate using the
two methods, in most cases within about $0.1$~eV.
Another indication of the quality of the method is the agreement between 
the gap obtained from total energy difference $\tilde{E}_g=E_S(+)+E_S(-)-2E_S(0)-2\delta_E$,
where $E_S(q)$ is the energy of the pristine supercell in charge state $q$ and
$\delta_E$ is the electrostatic correction of Ref.\onlinecite{komsa-PRL-110-095505},
and the Kohn-Sham gap. These are respectively $\tilde{E}_g=$1.64 and 1.87~eV for MoS$_2$ and WS$_2$,
and $E_g=$1.65 and 1.77~eV for MoS$_2$ and WS$_2$.

Adsorbed Li is a shallow donor with a small ionisation energy $<$0.1~eV both in MoS$_2$ and WS$_2$.
This is mainly due to two effects. First, the relaxation of Li in the positive charge state,
which is of the order of 30 meV and is a physical effect; second,
a spurious band filling effect,\cite{carvalho-PRB-80-195205,lany-PRB-78-235104} which are larger in WS$_2$ due to the greatest dispersion of the lowest conduction band.
The bandstructure shows inequivocally that Li$\rm _{ad}$ is a shallow donor.
In effect, it merely gives out an electron to the conduction band,
changing little the matrix bandstructure in the vicinity of the gap (Supplementary Figure 1).

Substitutional Br and Cl are shallow donors only above room temperature. 
They contribute with an additional electron to populate a perturbed conduction band state.
The shallowest of them is Br$\rm _S$, with a ionisation energy of about 0.1-0.2~eV
both in MoS$_2$ and WS$_2$ (Table~\ref{tab:levels}).
Even though this is higher than the ionisation energy of shallow dopants in bulk materials such as Si or GaAs,
it is lower than the dopant ionisation energies in layered BN.\cite{oba-PRB-81-075125}

Substitutional P is found to be a very shallow acceptor, with activation energy $\sim0.1$~eV in MoS$_2$,
and $<0.1$~eV in WS$_2$, comparable to the uncertainty of the calculation.
Si is also an acceptor, though deeper.

It is noticeable that ionisation energies in WS$_2$ are usually smaller, despite its
larger calculated bandgap, suggesting that this material is easier to dope.

%\section{Summary}

In summary, we have shown that it is possible to dope MoS$_2$ and WS$_2$ with electrons or
holes by chemical substitution at the S site or adsorption on the top of the layer.
Amongst the shallow donors, Li$\rm _{ad}$ has the lowest ionisation energy.
The donated electron is predominantly localized on the transition metal $d$ states.
However, Li diffuses extremely fast in most materials and therefore
is not a good choice for high temperature applications.
Besides, Br$\rm _S$ and Cl$\rm _S$ are also donors, but have a higher ionisation energy.
The higher temperature required to excite the carriers is a trade-off
for the higher temperature stability of the defects.

Phosphorus is a shallow acceptor with a very low 
ionisation energy, comparable to the uncertainty of the calculation.
The wavefunction of the unpaired hole state is a valence-band like state,
predominantly localized on the transition metal layer.
This suggests that the ionized P$\rm _S$ defect will be a weak scattering center.
The combination between the high stability of P and the fact that
it contributes with a very delocalized electron to the material, 
preserving the characteristics of a 2D electron gas,
indicate that its extrinsic region would extend up to much higher temperatures than for Si,
that readily becomes intrinsic at about 800~K (Fig.~\ref{fig:plot}).

These findings open the way to the control of the
conductivity type in these two materials, 
offering a way to use MoS$_2$ and WS$_2$ for transistor parts other than the channel,
or even to integrate different funtionalities in the same layer.
This seems extremely promising for the design of electronic and optoelectronic 
devices for high temperature operation. 

\section*{Acknowledgments}
We thank the financial support from NRF-CRP award "Novel 2D materials with tailored properties: beyond graphene" (R-144-000-295-281).
We thank RM Ribeiro for providing pseudopotentials.
The calculations were performed in the GRC high performance computing facilities.

\begin{table}[p]
\caption
{Formation energy of substitutional impurities ($E_f$)
along with adsorption energies.
All values are in eV and refer to the neutral charge state.  
\label{tab:Esub}}
\begin{ruledtabular}
\begin{tabular*}{\columnwidth}{lcccc}
&\multicolumn{2}{c}{MoS$_2$} &\multicolumn{2}{c}{WS$_2$}\\
\hline
Defect   &$E_f^{\rm S-poor}(D_{\rm S})$& $E_f(D_{\rm ad})$& $E_f^{\rm S-poor}(D_{\rm S})$& $E_f(D_{\rm ad})$\\
\hline
Br$_{S}$ &$-$1.0&  $-$0.7        &$-$0.3&$-$0.7\\
Cl$_{S}$ &$-$1.5&  $-$0.9        &$-$0.9&$-$0.9\\
Li$_{S}$ &$-$0.7&  $-$2.0        &$-$0.9&$-$1.5\\
P$_{S}$  &$-$2.9&  $-$0.7        &$-$2.7&$-$0.6\\
Si$_{S}$ &$-$2.6&  $-$1.6        &$-$2.0&$-$0.9\\
\end{tabular*}
\end{ruledtabular}
\end{table}

\begin{table}[p]
\caption
{Defect-related levels in MoS$_2$ and WS$_2$.
$E(-/0)$ is given relative to $E_v$ and $E(0/+)$ is given relative to $E_c$.
FEM and MM stand for Formation Energy Method and Marker Method, respectively (see text).
All values are in eV.
\label{tab:levels}}
\begin{ruledtabular}
\begin{tabular*}{\columnwidth}{lcccc}
\multicolumn{5}{c}{MoS$_2$} \\
\hline
& $(-/0)$ &$(-/0)$ & $(0/+)$ &$(0/+)$ \\
Method      & FEM & MM &FEM&MM\\
\hline
Br$_{S}$ & --  & --  &0.15 &0.22\\
Cl$_{S}$ & --  & --  &0.18 &0.27\\
Li$_{\rm ad}$ & --&--&$-$0.02&0.12\\
P$_{S}$  &0.11 & 0.06& --  &--\\
Si$_{S}$ &0.39 & 0.34& --  &--\\
\hline
\multicolumn{5}{c}{WS$_2$} \\
\hline
& $(-/0)$ &$(-/0)$ & $(0/+)$ &$(0/+)$ \\
Method      & FEM & MM &FEM&MM\\
\hline
Br$_{S}$ & --  & --     & 0.14&0.14\\
Cl$_{S}$ & --  &--      & 0.18&0.22\\
Li$_{\rm ad}$ & --&--&$-$0.36&$-$0.16\\
P$_{S}$  & 0.02&$-0.09$ & -- & --\\
Si$_{S}$ & 0.23& 0.12   & -- & --\\ 
\end{tabular*}
\end{ruledtabular}
\end{table}

\begin{thebibliography}{99}
\bibitem{note} The values of the correction for a defect in the middle of the slab, at the interface and on the surface were 0.19, 0.21 and 0.19 eV respectively for MoS$_2$ and 0.19, 0.21 and 0.16 eV respectively for WS$_2$.
\bibitem{footnote}{The value of $E_v$ obtained from the 
bandstructure of the pristine supercell is usually 
corrected by alignemento of the average electrostatic
potentials \cite{walle-jap-95-3851}. 
However, since in this case a great fraction of the supercell
was occupied by vaccuum in the alignment term was found to be negligible.}.
\end{thebibliography}
\end{document}